# High field magneto-transport in high mobility gated InSb/InAlSb quantum well heterostructures

A. M. Gilbertson[1,2], W. R. Branford[2], M. Fearn[1], L. Buckle[1] P. D. Buckle[1], T. Ashley and L. F. Cohen[2]

[1]*QinetiQ, St. Andrews Road, Malvern, Worcestershire, WR14 3PS, United Kingdom*
[2]*Blackett Laboratory, Imperial College London, Prince Consort Rd., London, SW7 2BZ, United Kingdom*

We present high field magneto-transport data from a range of 30nm wide InSb/InAlSb quantum wells. The low temperature carrier mobility of the samples studied ranged from 18.4 to 39.5 $m^2V^{-1}s^{-1}$ with carrier densities between $1.5 \times 10^{15}$ and $3.28 \times 10^{15}$ $m^{-2}$. Room temperature mobilities are reported in excess of 6 $m^2V^{-1}s^{-1}$. It is found that the Landau level broadening decreases with carrier density and beating patterns are observed in the magnetoresistance with non-zero node amplitudes in samples with the narrowest broadening despite the presence of a large *g*-factor. The beating is attributed to Rashba splitting phenomenon and Rashba coupling parameters are extracted from the difference in spin populations for a range of samples and gate biases. The influence of Landau level broadening and spin-dependent scattering rates on the observation of beating in the Shubnikov-de Haas oscillations is investigated by simulations of the magnetoconductance. Data with non-zero beat node amplitudes are accompanied by asymmetric peaks in the Fourier transform, which are successfully reproduced by introducing a spin-dependent broadening in the simulations. It is found that the low-energy (majority) spin up state suffers more scattering than the high-energy (minority) spin down state and that the absence of beating patterns in the majority of (lower density) samples can be attributed to the same effect when the magnitude of the level broadening is large.

## I. INTRODUCTION

InSb has been the subject of numerous experimental and theoretical studies. Characteristic features of the bulk crystal such as a narrow band gap and light effective mass, along with its heavy constituent atoms result in (*i*) an inherent large spin-orbit (SO) coupling arising from the Dresselhaus effect lifting the spin degeneracy in zero magnetic field [1] (*ii*) a large negative Landé *g*-factor ~ -50 (at the band edge) [2] and (*iii*) a high intrinsic carrier mobility [3]. These properties are present also in two-dimensional electron gases (2DEGs) formed in InSb quantum wells (QWs) which offer great potential for device applications. Advances in the growth of high quality InSb heterostructures have resulted in extrinsic carrier mobilities μ recently reported in excess of 5 $m^2V^{-1}s^{-1}$ at room temperature making InSb QWs particularly attractive for high-speed electronics (high electron mobility transistors) [4], ballistic transport devices and magnetic sensor applications such as non-magnetic read heads based on extraordinary magneto-resistance (EMR) [5].

In addition to more traditional charge based electronic devices, InSb QW heterostructures are considered to be promising candidates for spintronic applications such as the spin transistor proposed by Datta and Das [6] due to a large Rashba type SO coupling which provides an additional source of zero field spin splitting arising from the structural inversion asymmetry (SIA) in the heterostructure [7]. Such devices rely on the concept that the size of the Rashba coupling, parameterised by the coefficient $\alpha_R$, can be tuned via the application of an external electric field (i.e. a gate electrode). Although a large SO coupling results in a short spin lifetime ($\tau_s$ ~ 0.3ps at

300K in 20nm InSb/InAlSb QWs [8]) which reduces the spin diffusion length, this is mitigated somewhat by the small effective mass.

The two most common techniques for measuring the strength of the SO coupling are (*i*) in the observation and fitting of quantum interference corrections in the low field magnetoconductance to weak anti-localisation (WAL) theory [9] and (*ii*) in the analysis of beating patterns in the low field Shubnikov-de Haas (SdH) oscillations in the longitudinal resistivity $\rho_{xx}$ [10]. The accuracy and indeed the validity of the later approach can be somewhat controversial due to alternative explanations for the occurrence of beating [11] and the influence of Zeeman splitting [12]. Values obtained from this technique are generally larger that those obtained from WAL. Extensive WAL experiments have been performed on InSb thin films on GaAs (100) [13] and InSb/CdTe heterojunctions [14], providing unambiguous evidence for the presence of SO coupling in 2DEGs formed at the heterointerfaces. In contrast, only a small number of elegant measurements of the SO coupling have been made in InSb QWs; some of which are rather indirect and none using beating which has not been observed previously in InSb QWs. A large *g*-factor means that Zeeman splitting dominates the SdH oscillations at relatively small fields compared to other systems making the observation of beating patterns particularly challenging in narrow gap systems. Dedigama *et al.* recently reported the first observations of WAL in InSb QWs supporting the presence of large SO coupling, although only a preliminary empirical analysis was given [15]. Khodaparast *et al.* [16] studied spin splitting in asymmetric 30nm InSb/InAlSb QWs via electron spin resonance (ESR) in which a spin splitting was extrapolated to zero field. Assuming that the Rashba SO interaction was dominant the Rashba coupling parameter was estimated as $\alpha_R = 1.3 \times 10^{-11}$ eVm. A recent theoretical study of the SO coupling parameters in various InSb/InAlSb QW structures based on self-consistent band profile calculations and an eight-band k.p model [17] predicted smaller values of $\alpha_R$ in the range $2\text{-}7 \times 10^{-12}$ eVm and that the Dresselhaus contribution to spin splitting can be of significant and comparable value to that of the Rashba dependent on the details of the heterostructure [18]. Indeed, a recent study by Akabori *et al.* [19] demonstrated that the Dresselhaus SO interaction was dominant in a similar InGaSb/AlInSb QW sample. Clearly discrepancies exist between experiment and theory of the spin splitting phenomena in narrow gap systems, and the subject would benefit from a comprehensive investigation of samples with a range of carrier densities.

In this paper we present high field magneto-transport measurements on high mobility *n*-InSb/InAlSb QWs with varying carrier density and mobility as a function of temperature and gate bias. Similar samples were previously measured where a preliminary analysis was made in an attempt to extract information on the Rashba spin splitting [20]. Here we perform a comprehensive study on a wider range of samples. From analysis of the data and with the use of magnetoconductance simulations, we propose that the direct measurement of SO coupling in InSb QWs is usually elusive due to the combination of large inhomogeneous, spin-dependent broadening combined with a large Zeeman spin splitting. The paper is organised in the following way. In section II a description of the experiment and samples is given. In section III the experimental results and analysis are presented. Finally, in section IV some conclusions are drawn.

## II. EXPERIMENT

Samples were grown by solid source molecular beam epitaxy (MBE) onto semi-insulating GaAs (001) substrates. A schematic view of the layer structure along with a typical band profile generated from a Schrodinger-Poisson model near the surface are shown in Fig. 1. In this calculation mid-gap pinning of the Fermi energy at the surface boundary was assumed. In growth order, the heterostructure consists of an accommodation layer, a 3um intentionally undoped $In_{0.9}Al_{0.1}Sb$ buffer layer, a 30nm strained InSb QW and a 50nm $In_{0.85}Al_{0.15}Sb$ upper barrier forming a type-I heterostructure confining both electrons and holes. The upper barrier was δ-doped with Te, separated from the QW by an undoped spacer layer of thickness $S = 20$nm. As seen in Fig. 1 the resulting QW is asymmetric both in physical barrier composition and electrostatic confining potential in the growth direction. Low field electron transport studies in these heterostructures have recently been performed, indicating that carrier mobility in these remote doped wide well structures is dominated by remote ionised impurity scattering (RIIS) at low temperatures [21].

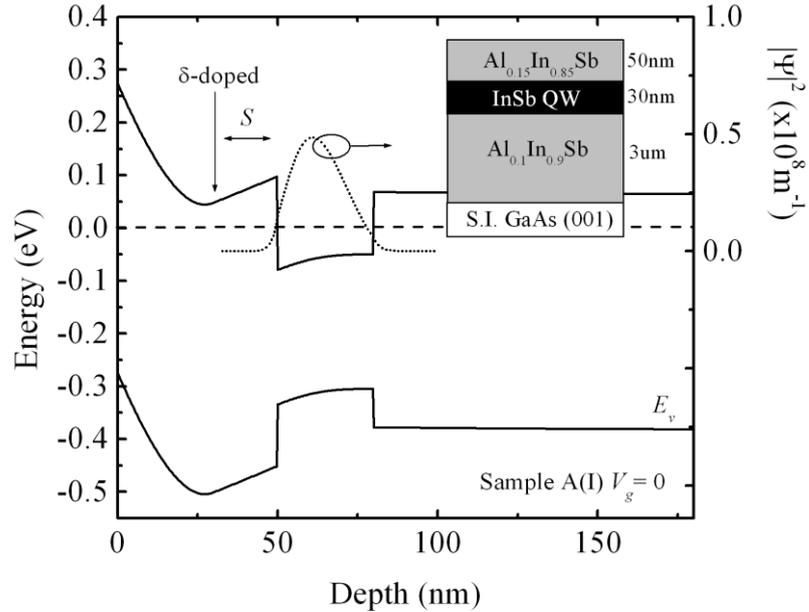

FIG. 1. Schrodinger-Poisson solution for sample A(I) at zero gate bias and 10K in the vicinity of the QW showing the band profile (solid back line) and the single occupied ground state (dotted line) beneath the Fermi energy $E_f = 0$ meV. The ground state probability density function is also shown indicated by the solid red line. The position of the Te δ–layer is indicated where $S$ is the un-doped spacer layer thickness. (Inset) A schematic view of the layer structure.

Magneto-transport measurements were performed using conventional 40μm wide gated Hall bridges fabricated using optical lithography and wet etching with voltage probes separated by 200μm. Shallow contact techniques were employed to form the Ohmic contacts and ensure that transport is via the 2D channel only. Ti/Au top gate electrodes were evaporated onto an insulating $SiO_2$ dielectric layer (see Table 1) which covered the sample. Note that voltage probes in these devices are located sufficiently away from the current contacts so that geometric effects can be ignored [22].

TABLE 1. Sample parameters $n_{2D}$ and µ at 2K and 290K (at zero gate bias) along with nominal gate oxide thickness.

| Sample | µ (m$^2$V$^{-1}$s$^{-1}$) 2K (290K) | $n_{2D}$ (m$^{-2}$) 2K (290K) | Gate oxide (nm) |
|---|---|---|---|
| Sample A(I) | 27.2 (6.78) | 2.32 x10$^{15}$ (3.29 x10$^{15}$) | 50 |
| Sample A(II) | 26.13 | 2.51x10$^{15}$ | 50 |
| Sample B(I) | 39.5 | 3.28 x10$^{15}$ | 150 |
| Sample B(II) | 39.0 | 3.21 x10$^{15}$ | 150 |
| Sample C | 18.36 (5.07) | 1.50 x10$^{15}$ (4.12 x10$^{15}$) | 50 |

The devices were measured in a cryogen free magnet system enabling measurements to be performed over a magnetic field range of -7.5T < B < 7.5T and temperatures down to 2K. Longitudinal and Hall resistivities $\rho_{xx}$ and $\rho_{xy}$ were measured with magnetic field applied perpendicular to the plane of the 2DEG using a low-frequency lock-in technique at drive currents of less than 500nA (the observed SdH oscillations were strongly dampened at drive currents >1µA due to Joule heating).

### III. RESULTS AND DISCUSSION

Five samples were investigated that were fabricated from three different wafers. These are labelled sample A (I and II), sample B (I and II) and sample C. To characterise the 2DEGs, the sheet carrier density $n_{2D}$ at zero gate bias was determined both from low-field Hall Effect measurements and from the SdH fundamental frequency which agreed to within 2% indicating that no parallel conduction paths are present. These values and the associated carrier mobilities, µ, are listed in Table 1 for each sample at 2K and 290K. A small variation in carrier density was observed between devices from the same wafer due to the sensitivity of the 2DEGs to the presence of (spatially non-uniform) surface states at the dielectric/InAlSb interface. It is noteworthy that the samples investigated here exhibit the highest low temperature mobilities reported in the InSb QW system and the highest RT mobilities in all III-V QW systems reported.

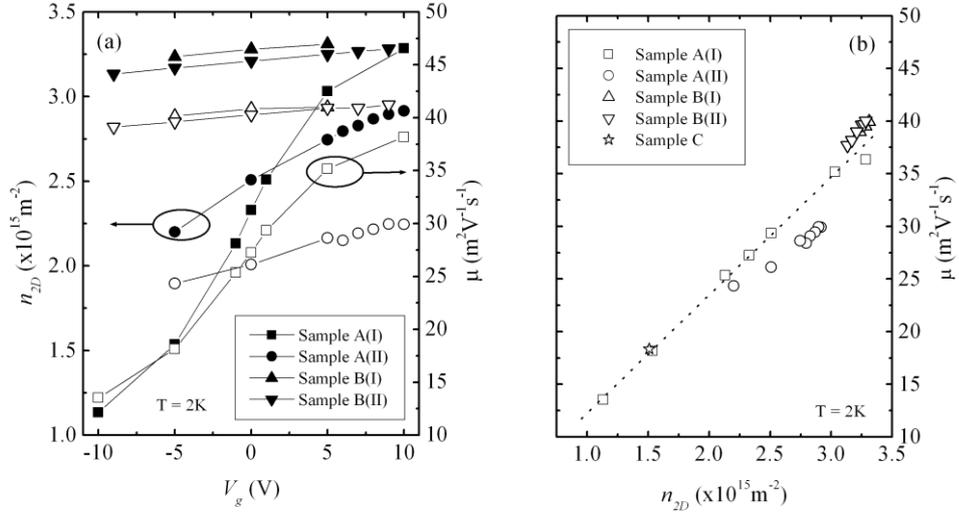

FIG. 2. (a) Sheet density $n_{2D}$ (closed symbols, left axis) and mobility μ (open symbols, right axis) for samples A(I) (squares), A(II) (circles), B(I) (upward triangle) and B(II) (downward triangle) as a function of gate bias $V_g$ at 2K. (b) The 2K mobility as a function of carrier density for all samples and gate biases showing a linear relationship. The dashed line is a guide to the eye.

Using the top gate electrode we were able to modulate $n_{2D}$ and μ in samples A(I), A(II), B(I) and B(II) over a range of values. Data for $n_{2D}$ (closed symbols) and μ (open symbols) as a function of gate bias $V_g$ obtained at 2K is presented in Fig. 2(a). Due to difficulties in producing reliable low leakage gate dielectrics, samples B(I) and B(II) were fabricated with a nominally thicker dielectric layer (see Table 1) which is reflected in the smaller modulation of $n_{2D}$ shown in Fig. 2(a). The gate electrode for sample C did not function due to excessive leakage current and so we focus our discussion on the remaining four samples. For each device as $n_{2D}$ is increased, μ increases steadily. This behaviour is shown more clearly in Fig. 2(b) and is typical for modulation doped heterostructures whereby the increasing Fermi velocity in the 2DEG reduces the effectiveness of the Coulomb scattering from remote ionised impurities and subsequently increases the momentum scattering lifetime (related to mobility by $\tau_p = m^*\mu/e$, where $m^*$ is the effective mass and $e$ the electron charge [23]).

Fig. 3(a) shows typical low temperature recordings of the longitudinal resistivity $\rho_{xx}$ and Hall resistance $\rho_{xy}$ from sample B(I) in the range 2K to 20K ($V_g = 0$). At quantising magnetic fields μB >> 1 Landau levels (LLs) are resolved in the density of states (DoS) and plateaus emerge in the Hall resistance, quantised to values of $\rho_{xy} = h/ie^2$ (with $i = 1, 2,..$). The plateaus in $\rho_{xy}$ are accompanied by minima in the SdH oscillations in $\rho_{xx}$ corresponding to when the Fermi energy lies between two LLs.

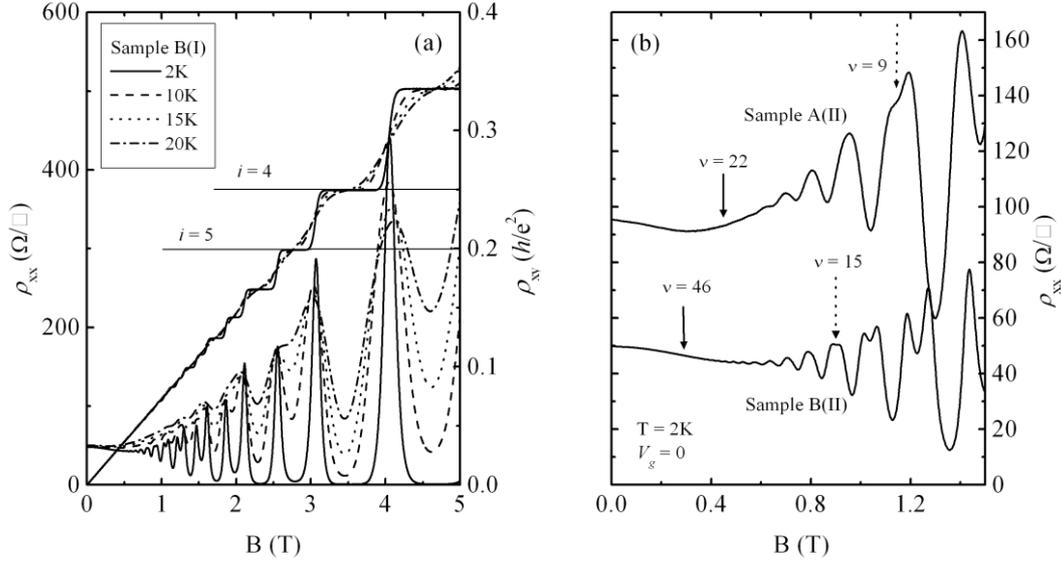

FIG. 3. (a) Longitudinal $\rho_{xx}$ (left axis) and transverse $\rho_{xy}$ (right axis) resistivity measured at various temperatures as a function of magnetic field B for sample B(I) ($V_g = 0$) demonstrating single period SdH oscillations and integer quantum Hall effect. (b) Low field region of $\rho_{xx}(B)$ for samples A(II) (upper trace) and B(II) (lower trace) indicating the onset of SdH oscillations (solid arrows) and the emergence of Zeeman splitting (dotted arrows) at odd filling factors as high as $\nu = 15$.

Clear single-period oscillations are observed for all samples and gate biases indicative of single subband occupation which is supported by Schrodinger–Poisson solutions of the band profiles (see Fig. 1). In Fig. 3(b) we show 2K magnetoresistance data for samples A(II) (upper trace) and B(II) (lower trace) at 2K where SdH oscillations are resolved at filling factors ($\nu = n_{2D}h/eB$) up to $\nu = 46$ in the higher mobility sample as indicated by the solid arrows. Also indicated by the dotted arrows is the emergence of Zeeman splitting at odd filling factors as high as $\nu = 15$, demonstrating the presence of a large $g$-factor. Reducing the temperature below 5K gave no significant improvement in the resolution of the low field SdH oscillations (not shown), indicating that at 2K the SdH oscillations are limited by inhomogeneous LL broadening rather than thermal broadening. This is in good agreement with measurements made on similar samples down to lower temperatures (300mK) [ref Nicholas].

From examination of Figs. 3(a) and 3(b) it can be seen that there is a distinct non-oscillatory background magnetoresistance present in our samples which is temperature dependent. At low fields $\rho_{xx}$ contains at first negative and then a positive magneto-resistance, which becomes approximately linear at high fields. The low field region B < 0.4T depicted in Fig. 3(b) is consistent with the effects of electron-electron interactions in the presence of Zeeman spitting as described by Lee and Ramakrishnan [24], although, this mechanism will not be examined here. The high field quasi-linear magneto-resistance has previously been observed in InSb epilayers and is attributed to the intrinsic magnetoresistance originating from sample inhomogenieties [22,25].

## A. Estimation of Landau level broadening

As indicated by the position of the solid arrows in Fig. 3(b), the extent of the low field SdH oscillations of interest varies between samples. This is strongly influenced by the broadening of the LIs, $\Gamma$, and a more quantitative examination is crucial. Under the assumption that the broadening has no significant thermal contribution, a simple estimate for $\Gamma$ is made from the critical field at which SdH oscillations become resolved, denoted here by $B_{SdH}$. Oscillations in $\rho_{xx}$ are a manifestation of the oscillations in the DoS and so it is reasonable to assume that these will become resolved when the cyclotron energy exceeds the level broadening, then the broadening is given simply by $\Gamma = \frac{\hbar e B_{SdH}}{m^*}$. Due to its narrow band gap, the conduction band of InSb is highly non-parabolic and the mass becomes energy dependent. It is therefore necessary to consider these effects on $m^*$ in order to estimate $\Gamma$. Within the six-band Kane model [26] the conduction band (near $k = 0$) can be described by the dispersion relation $E(1 + E/E_g) = \hbar^2 k^2 / 2m^*_{cb}$, where $E$ is the electron energy, $k$ the wavevector, $E_g$ is the band gap and $m^*_{cb}$ is the effective mass at the conduction band edge. The effective mass is related to the first derivative of the dispersion relation with respect to wave vector and is given by [27],

$$m^*(E) = m^*_{cb}\left(1 + \frac{2E}{E_g}\right) \quad (1)$$

Since we are interested in the conduction at the Fermi energy, $E = E_F$ and using $k_F = (2\pi n_{2D})^{1./2}$ we have:

$$E_F = \left(\frac{E_g^2}{4} + \frac{E_g \pi \hbar^2 n_{2D}}{m^*_{cb}}\right)^{\frac{1}{2}} - \frac{E_g}{2}. \quad (2)$$

This description of the effective mass agrees well with recent experimental data obtained from the temperature dependence of SdH oscillations in similar InSb QWs [ref Nicholas]. Using Eqs. 1 and 2 with the parameters $m_{cb}^* = 0.014 m_0$ and $E_g = 0.255$ eV (taking into account the effect of strain) [3] we calculate appropriate values for $m^*$ which are then used in the estimation of $\Gamma$. Careful examination of both first and second derivatives of $\rho_{xx}(B)$ was required in order to determine $B_{SdH}$. The results of this analysis for each sample at zero gate bias are listed in Table 1. The same treatment was repeated for each gate bias measured and the resulting values for $\Gamma$ are plotted against carrier density in Fig. 4.

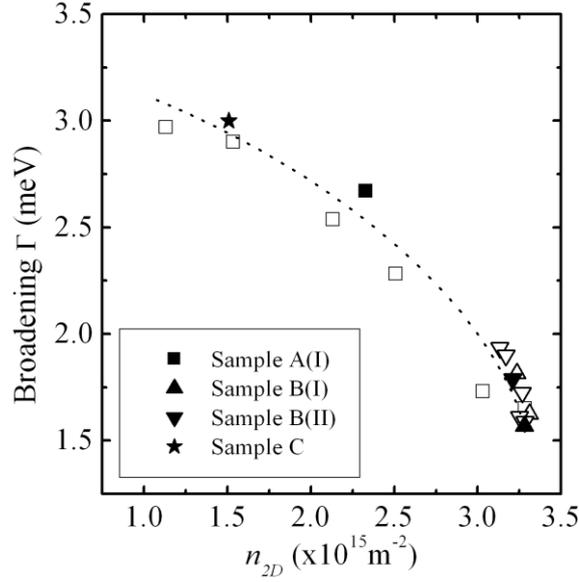

FIG. 4. Landau Level broadening parameter $\Gamma$ for each sample at different gate biases as a function of carrier density as determined from the critical SdH field $B_{SdH}$. Closed symbols represent data at zero gate bias. The dashed line is a guide to the eye.

A relationship is found between $\Gamma$ and $n_{2D}$ such that data from different samples appear to fall close to a single line. It should be emphasised that although this simple approach has the advantage that it makes no assumptions of the scattering potential, in practice it is limited by the signal-to-noise ratio (SNR) of the experiment which can limit the resolution of $B_{SdH}$ leading to an overestimation of $\Gamma$. This was apparent in sample A(II) where the SNR in the raw data was significantly lower than in the measurements of other samples.

The magnitude of $\Gamma$ in these samples ranging from 1.5-3meV is surprisingly large compared to typical values extracted from GaAs/AlGaAs and InAs/GaSb systems of $\Gamma$~0.26meV [28] and $\Gamma$~0.4-1.5meV [29,30] respectively. The effect of the level broadening on the extraction of the spin splitting is discussed later in the section. It is somewhat counterintuitive that $\Gamma$ is large and yet the mobility $\mu$ is high, suggesting that the scattering processes that influence $\Gamma$ do not adversely effect $\mu$. The nature of the broadening depends strongly on the range of the scattering potentials involved [31]. Since these structures have been shown to be limited by RIIS at low temperatures, it is not unreasonable to suggest that the large broadening in these samples results from the long range nature of the scattering potential associated with remote doping. In this regime, $\Gamma$ is susceptible to, and determined by inhomogeneities in the local potential energy felt by the carriers. Such inhomogeneities may result from spatial variations in well width and/or interface roughness in the sample and perhaps reflects the difficulty in the growth of high quality InSb heterostructures on highly mismatched GaAs substrates. However, it is interesting to note that estimating $\Gamma$ from data taken from a similar InSb QW sample grown from a different MBE source appears to show similar levels of broadening [16] to those found here. We point out that this conjecture is clearly not universal for remote doped heterostructures, e.g. a narrow broadening of ~0.6meV was found in a similar narrow gap InGaSb/AlInSb structure with a large 50nm spacer layer in Ref. [19]. In this case the mobility was relatively small compared to samples studied here and it is plausible that alloy scattering (short range) in the InGaSb channel in their sample may have influenced the transport.

## B. Spin splitting analysis

Various authors have reported beating in the low field SdH oscillations in the InAs [30,32,33] and InGaAs [10,34-36] systems which is assigned to SO splitting of the conduction band. Beating patterns are thought to arise from the participation of two sets of SdH oscillations with similar amplitudes differing slightly in frequency analogous to optical beating. This corresponds to the presence of two types of carriers with similar densities and effectives masses and is thus attributed to the spin splitting of the ground state rather than the occupation of two 2D subbands. This allows for extraction of the total spin splitting or if dominant, the Rashba coefficient $\alpha_R$ from either the field dependence of the beat node positions [10] (if more than three nodes are observed), or from the difference in carrier densities of the two spin populations $\Delta n = n_1 - n_2$ determined from a fast Fourier transform (FFT) of the low field $\rho_{xx}$ data [12]. The observation of beating in $\rho_{xx}$ has not been made in the InSb QW system to date.

Measurements from all samples at each gate bias show no *obvious* beating in the low field $\rho_{xx}$ data as previously reported in other systems. This can be seen in the data of Fig. 3(b) for samples A(II) and B(II) and Fig. 5(a) for B(I). However, careful inspection of the first and second derivative of the $\rho_{xx}$ data with respect to B reveals a weak modulation in the SdH oscillation amplitude, far from the onset of resolved Zeeman splitting. This is demonstrated in Fig. 5(b) which shows a beating pattern in the second derivative of the same data presented in the upper panel plotted against inverse field (data has been smoothed by three-point adjacent averaging). The positions of the beat nodes for this data are indicated by arrows.

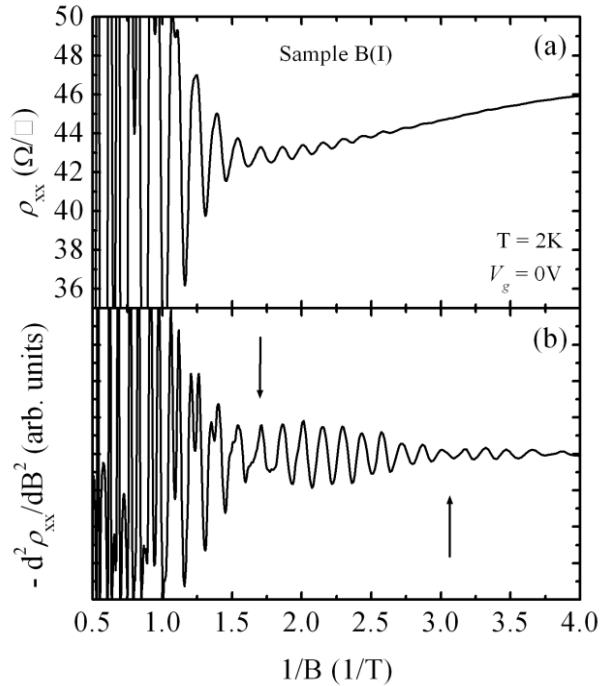

FIG. 5. (a) (Upper panel) Low field region of $\rho_{xx}$ plotted against inverse field 1/B for sample B(I) at $V_g$ = 0V and T = 2K. (b) (lower panel) Second derivative of the same data from sample B(I) showing the resolution of a clear beating pattern. The arrows show the positions of the beat nodes.

Weak beating is exhibited for all gate biases in samples B(I), B(II) and in the $Vg$ = 10V data only in sample A(I) where the broadening is small. It is not observed in samples A(II) or C in which the broadening is large.

It is important to rule out erroneous identifications of zero-field spin splitting from the observation of beating. It was shown by Rowe *et al.* [11] in the InAs/GaSb system that beating patterns can arise from the mixing of the SdH series from the ground-state subband and a magneto-intersubband (MIS) series, which is unrelated to zero field spin splitting. MIS scattering can only occur when the second subband becomes occupied. In our samples where beating is observed, no evidence for second subband occupation is found in either the FFT spectra or in the gate dependence of the carrier mobility which is supported by self-consistent band profile calculations (see Fig. 1). We can also exclude that the beating is a result of inhomogeneous carrier density since the beating patterns are present in three different samples and the length scale of the Hall bridges is small (<200μm). Thus we can attribute the observed beating to spin splitting phenomena.

We have recently calculated the SO coupling parameters in our material system for various carrier densities using self consistent calculations and results from the k.p method [18]. While the strength of the *k*-linear (*β*) and *k*-cubic (*γ*) Dresselhaus couplings are high in these wide well structures ($\beta \sim 3\times10^{-12}$eVm and $\gamma \sim 430$eVÅ$^3$ at $3\times10^{15}$m$^{-2}$) and contribute significantly to the spin splitting at low densities where the Rashba effect is suppressed, their significance rapidly decreases with density as the Rashba coupling is enhanced due to increasing electrostatic asymmetry across the QW [18]. Thus, at the higher carrier densities where beating is observed in our samples, the Rashba effect can be assumed to be the dominant mechanism.

For data where beating is observed, performing the FFT reveals a double peaked structure from which the carrier densities of the two spin populations $n_1$ and $n_2$ can be extracted according to $n_{1,2} = f_{1,2}e/h$. Here $f_{1,2}$ is the FFT frequency (in Tesla) of the two peaks. An example of the typical FFT spectrum is displayed in the inset of Fig. 6 showing the double peak structure from which $n_1$ and $n_2$ are determined. The asymmetry in the peak amplitudes is observed in all cases and is discussed in a later section. The relatively poor resolution of the FFT spectra is due to the small number of oscillations in the low field window of these low density samples. This introduces uncertainties in the peak positions $f_{1,2}$ which are taken into account in the extraction of $\Delta n$.

With only one or two beat nodes distinguishable from our data the Rashba parameter can not be determined from the positions of the beat nodes [10]. Instead, the Rashba SO coupling parameter can be determined from the difference in the spin populations (from the FFT spectra) by the expression given by Engels *et al.* [12]:

$$\alpha_R = \frac{\Delta n \hbar^2}{m^*} \sqrt{\frac{\pi}{2(n_{2D} - \Delta n)}} \,. \tag{3}$$

This expression is derived from the parabolic energy dispersion in the presence of Rashba splitting [7] which leads to a spin-dependent DoS in zero field. It can be shown that incorporating the effects of band non-parabolicity using the effective two-band model in the derivation of Eq. 3 yields the same result. For consistency, we use the appropriate effective mass for each value of $n_{2D}$ as described earlier. The results of this analysis for data sets where splitting in the FFT spectra was distinguishable are presented in Fig. 6 as a function of carrier density.

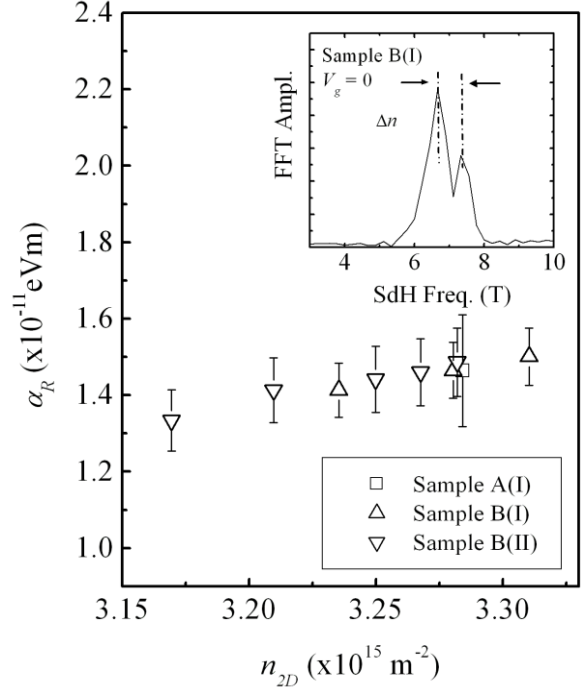

FIG. 6. Values for the Rashba coefficient $\alpha_R$ obtained from experimentally extracted difference in spin populations $\Delta n$ as a function of carrier density $n_{2D}$ for samples A(I), B(I) and B(II). Error bars originate from the uncertainty in FFT peak positions. (Inset) Typical FFT amplitude spectrum from B(I) at $V_g = 0$V from which $\Delta n$ is extracted.

The value of the Rashba parameter extracted from samples B(I) and B(II) increases monotonically with carrier density. This is in contrast to the results of Nitta *et al.* [34] Engels *et al* [12], Schapers *et al.* [36] and Hu *et al.* [37] obtained in the InGaAs system who reported values of $\alpha_R$ which decreased with carrier density. These results were obtained from top (front) gated structures where the doping layer is positioned below the QW. Both these dependences are fully accounted for by theoretical treatments that take into account finite barriers [17,18]. In this theory it is observed that the Rashba parameter is determined predominantly by the difference of the electron probability density function at the upper and lower QW interfaces, such that,

$$\alpha_R \propto <E_z> + [A\psi_{UI}^2 - B\psi_{LI}^2], \quad (4)$$

where the first term $<E_z>$ represents the electric field averaged over the ground state wave function and the second term represents the difference in the probability density functions at the upper $\Psi_{UI}^2$ and lower $\Psi_{LI}^2$ interfaces weighted by coefficients $A$ and $B$ which contain the material parameters and band offsets [17]. As shown in Fig. 1, in our asymmetric structures doped above the QW, the ground state wave function is weighted towards the upper interface. Application of positive gate bias lowers the potential at the left hand boundary ($z = 0$) which has the twofold effect of increasing the electric field across the QW (first term in Eq. 4) and skewing the wavefunction further towards the upper interface (second term in Eq. 4), thus increasing the Rashba parameter and accounting for the trend observed in our samples. Since in structures doped below the QW the direction of the electric field is reversed, it follows that the opposite dependence of the Rashba parameter on carrier density is observed. This concept is further supported by the results of Grundler, who, with the independent use of both a top and back-gate, could observe both dependencies of $\alpha_R$ on $n_{2D}$ in an InAs 2DEG doped below the QW [33] (using a back gate on a sample doped below the QW is equivalent to our experiment).

A striking feature of Fig. 6 is that data from all three samples appear to lie on a common line. This trend is in agreement with calculations of SO parameters in InSb QW structures [18] where varying the electrostatic potential in the heterostructure via the doping density and spacer thickness (or indeed gate bias), demonstrated that when plotted against a common variable such as carrier density the data fall onto a common curve.

We can make some comment on the magnitude of $\alpha_R$ extracted from our samples. Previously in the InSb QW system, the Rashba parameter has been extracted from ESR measurements in tilted fields by Khodaparest *et al.* [16] giving $\alpha_R = 1.33 \times 10^{-11}$eVm for a 30nm QW. This agrees well with our extracted values of $\alpha_R$. However, these values obtained experimentally are considerably larger than theoretical calculations of the Rashba parameter in these heterostructures which predict $\alpha_R = 6.5 \times 10^{-12}$eVm for a 30nm QW at $n_{2D} \sim 3 \times 10^{15}$m$^{-2}$ i.e. a factor of two smaller than that extracted experimentally. We note that the expression used in Eq. 3 has the disadvantage of being derived from the DoS in zero magnetic field, while values of $\Delta n$ used to calculate $\alpha_R$ are determined from measurements in non-zero fields. Consequently, the contribution from the Zeeman term is neglected and the accuracy of the magnitude of $\alpha_R$ holds some ambiguity. This may be of particular significance in narrow gap systems where $g^*$ is large. In addition, discrepancies between theory and experiment may result from the presence of many body effects such as the exchange interaction which are not included in the k.p approach [ref]. It was shown that the interaction effect in the presence of SO coupling can lower the energy levels of a system and enhance the spin splitting [38]. Large exchange enhancement of the *g*-factor in the InSb QW system has recently been demonstrated [ref nicholas] and so it may be important here. Previous studies of spin splitting using high field magneto-transport measurements in InAs and InGaAs QWs have been performed on samples with narrow QWs and carrier densities in excess of $1 \times 10^{16}$m$^{-2}$. Due to the reduced density of states in the InSb system and the requirement of single subband occupation in these wide QWs, the carrier density in our sample is significantly lower, limiting direct comparisons in these studies to just a few cases. Recently, Guzenko *et al.* studied the Rashba splitting in a low doped InGaAs/InP sample by analysis of WAL and beat node positions which were found to be complimentary [35]. In this study the carrier density of the 2DEG was varied over the range 1.3 to $7.3 \times 10^{15}$m$^{-2}$ and a value of $\alpha_R \sim 6.5 \times 10^{-12}$eVm was extracted at a carrier density of $n_{2D} \sim 3.1 \times 10^{15}$m$^{-2}$ from beating analysis. This value is approximately half that extracted from our samples which is consistent with the trend in the literature and the expectation that the Rashba parameter scales inversely with band gap [39].

## C. The influence of level broadening on beating patterns

We speculate that the absence of beating in samples A(II), C and A(I) for $V_g < 10$V can be attributed to the combination of *large broadening $\Gamma$ and a large Zeeman splitting* which limits the field range $B_{SdH} < B < B_Z$ over which the effects of SO splitting are observable. Here $B_Z$ is the field at which Zeeman splitting is resolved, i.e. for $B > B_Z$ spin splitting in the 2DEG is dominated by the Zeeman effect [40]. The number of oscillations within this field range depends on both $\Gamma$ and the carrier density $n_{2D}$ and from this assertion it follows that only in samples with the greatest number of oscillations are the effects of beating detectable (see Fig. 4).

This conjecture may be quantified by simulations of the SdH oscillations in the presence of Rashba SO splitting. In this analysis, we consider the result for the Landau level energy spectrum in the presence of Rashba splitting $E_{n\pm}$ for spin up (+) and spin down (-) given analytically by [41]:

$$E_{n\pm} = \hbar\omega_c\left(n + \frac{1}{2} \pm \frac{1}{2}\right)$$
$$\mp \frac{1}{2}(\hbar\omega_c - g^*\mu_B B)\sqrt{1 + \frac{8\alpha_R^2}{(\hbar\omega_c - g^*\mu_B B)^2}\frac{eB}{\hbar}\left(n + \frac{1}{2} \pm \frac{1}{2}\right)} \quad (5)$$

where $n = 1, 2,..$ is the Landau level index and $\omega_c = eB/m^*$ is the cyclotron frequency. Following the result of Gerhardts [42] the DoS takes on the Gaussian form and the magneto conductance at $T = 0$ K can be given by

$$\sigma_{xx}(B) = \frac{e^2}{2\pi\hbar}\sum_{n\pm}(n + 1/2)\exp\left(-\frac{(E_F - E_{n\pm})^2}{\Gamma_\pm^2}\right). \quad (6)$$

Here we denote $\Gamma_\pm$ as the broadening of spin up (+) and spin down (-) Landau levels respectively. To perform the simulations, the Fermi energy $E_F$ is first calculated at each field by solving the integral of the DoS multiplied by the Fermi distribution function in order to achieve the desired carrier density. The resistivity is obtained in the usual manner through inverting the conductivity tensor,

$$\rho_{xx} = \frac{\sigma_{xx}}{(\sigma_{xx}^2 + \sigma_{xy}^2)}. \quad (7)$$

where we use the classic expression for the Hall conductivity $\sigma_{xy} = -en_{2D}/B$ which is valid in low fields as done by previous authors [30,34]. $\Gamma$ is taken to be field independent.

Initially we set $\Gamma_+ = \Gamma_- = \Gamma$ and use combinations of parameters $n_{2D}$ and $\Gamma$ according to the results presented in Fig. 4 to simulate data from samples with narrow broadening where beating is observed in our experimental data and samples with larger broadening where it is not. The simulation with parameters $n_{2D} = 3.3 \times 10^{15}$m$^{-2}$, $\Gamma = 1.6$meV, $\alpha_R = 1.3 \times 10^{-11}$eVm and *g*-factor $g^* = -30$ is shown in the lower trace in Fig. 7 which exhibits a pronounced beating pattern. In contrast, we see that the simulation with parameters $n_{2D} = 2.5 \times 10^{15}$m$^{-2}$ and $\Gamma = 2.5$meV (using the same spin splitting parameters for consistency) shown by the upper trace in Fig. 7 shows no discernible beating pattern.

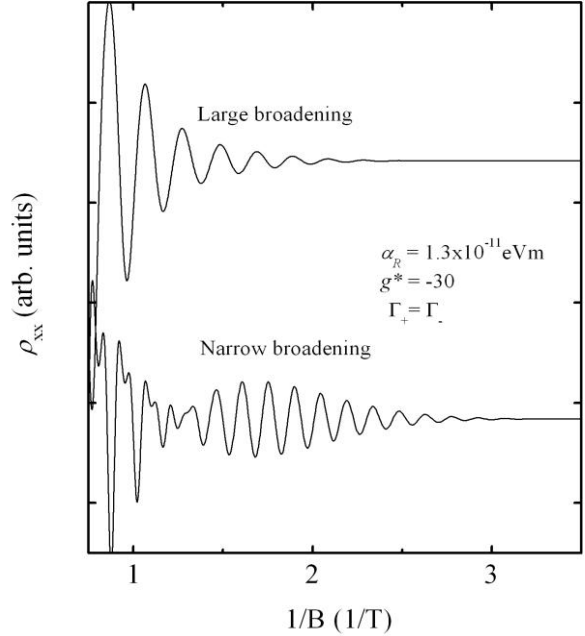

FIG. 7. Numerical simulations of the magneto resistance $\rho_{xx}$ with input parameters $\Gamma_+ = \Gamma_- = 1.6$meV and $n_{2D} = 3.3 \times 10^{15}$m$^{-2}$ (lower trace) and $\Gamma_+ = \Gamma_- = 2.5$meV and $n_{2D} = 2.5 \times 10^{15}$m$^{-2}$ (upper trace) according to data from Fig. 4, demonstrating the disappearance of beating patterns at larger broadening. Both simulations have a Rashba parameter of $\alpha_R = 1.3 \times 10^{-11}$eVm.

The disappearance of beating with large broadening in these simulations is consistent with the observations made in our samples and provides at least a semi-quantitative basis for interpreting the absence of beating in the majority of our samples (although it is expected that the Rashba parameter will be smaller for lower carrier density, which would reduce the beating pattern further). We note however, that in addition to the influence of broadening, due to competing spin splitting mechanisms which dominate in different regimes [18], the Dresselhaus splitting may not be negligible at lower carrier densities and may also influence the observed beating patterns. It is worth commenting that Brosig *et al.* [29] also reported the absence of beating in high quality InAs/AlSb and InAs/AlGaSb QWs over a range of carrier densities. In their samples however, SdH oscillations were resolved at fields as low as B ~ 0.15T with a narrow broadening of $\Gamma$ ~ 0.4meV and so the absence of beating in their samples can not be attributed to the same broadening mechanism.

It is interesting that the numerical simulation of a narrow broadened sample (lower trace Fig. 7) with the experimentally extracted Rashba parameter $\alpha_R$ ~ 1.3×10$^{-11}$eVm does not reproduce well the features in the experimental data (shown in Fig. 5(a)). The stronger beating pattern in the simulation suggests that the Rashba parameter used is larger than that in our samples, and the zero amplitude beat node is not observed in the data from our samples. In fact, a better agreement is found with the experimental data if smaller values of $\alpha_R$ are used.

## 1. Spin-dependent scattering rate

The non-zero beat node amplitude in our samples indicates that the SdH series originating from the two spin subbands oscillate at the Fermi energy with different amplitudes. This conjecture is strongly supported by the unequal amplitudes of the spin-split peak in the FFT spectra (see inset Fig. 6). The observation of non-zero beat node amplitude has also been made by Lou *et al.* [30] in a 10nm InAs QW which was qualitatively interpreted by introducing the concept of a spin-dependent scattering process which suppresses the oscillation amplitude of one spin more than the other (although the nature of the mechanism was not discussed). This interpretation is based on the understanding that at low temperatures the SdH amplitude in the low field region is predominantly determined by the single-particle relaxation time $\tau$ [43], which in remote doped structures is typically an order of magnitude smaller than the momentum scattering time [44]. Thus, in this interpretation there is a different scattering time $\tau_{\pm}$ associated with each spin. Importantly, it was shown by Lou *et al.* that the disparity between the two scattering times was proportional to the spin splitting in the system, be it from the SO interaction or external field, and thus the appearance of such features in our samples is consistent with the presence of a large spin splitting. Experimentally, $\tau$ is commonly extracted from the field dependence of the oscillation amplitude $\Delta\rho_{xx} \propto \exp(-\pi/\omega_c\tau)$ [44,45], however, this is difficult when the two sets of oscillations are superimposed in the low field region.

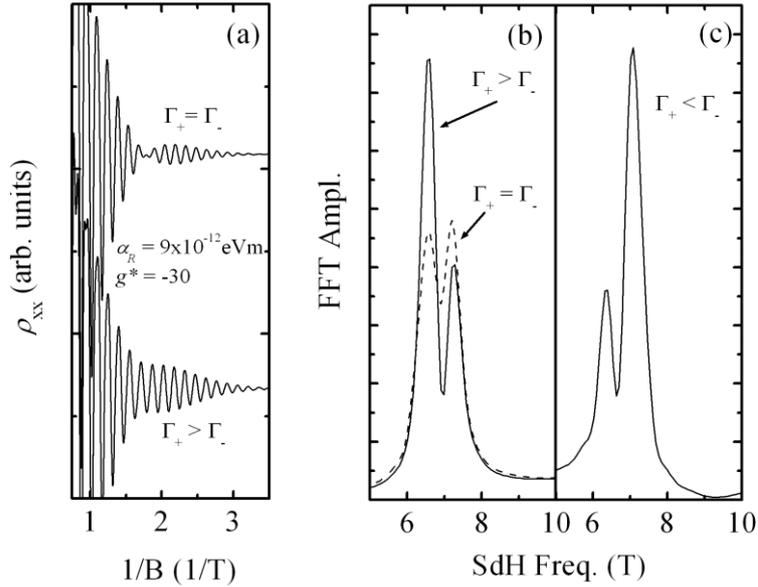

FIG. 8. (a) Simulations of $\rho_{xx}$ with $\alpha_R = 9\times10^{-12}$eVm, $n_{2D} = 3.3\times10^{15}$m$^{-2}$ and $\Gamma_+ = \Gamma_- = 1.6$meV (upper trace) compared to the results when using a spin dependent broadening $\Gamma_+ = 1.6$meV and $\Gamma_- = 1.4$meV. (b) Corresponding FFT spectra of the numerical simulations shown in (a) showing an asymmetry in the peak amplitudes with spin dependent broadening (solid line) compared to a symmetric structure with equal broadening for both spins (dashed line). (c) FFT spectrum from a simulation when the spin dependent broadening parameters are switched demonstrating the opposite asymmetry in peak amplitude.

To qualitatively explore this concept, we note that the broadening of the Landau levels is related to the single-particle relaxation rate by $\Gamma \propto \hbar\tau^{-\eta}$ where generally $\eta = 1$ or $\eta = ½$ dependent on the nature of the scattering [31]. The effects of spin-dependent scattering rates on the SdH oscillations can be incorporated into the numerical simulations by introducing a spin-dependent level broadening in Eq. 6 i.e. $\Gamma_+ \neq \Gamma_-$. The effect of spin-dependent broadening on the SdH oscillations is demonstrated in Fig. 8(a) for the case $\Gamma_+ > \Gamma_-$ with $\Gamma_+ = 1.6$meV and $\Gamma_- = 1.4$meV (lower trace) compared to equal broadening $\Gamma_+ = \Gamma_- = 1.6$meV (upper trace). Here we have used a smaller Rashba parameter of $\alpha_R = 9 \times 10^{-12}$eVm which gives closer resemblance to the experimental data. It can be seen that a spin-dependent broadening can indeed produce a non-zero beat node amplitude. The corresponding FFT spectra of the simulations with $\Gamma_+ > \Gamma_-$ and $\Gamma_+ = \Gamma_-$ are shown in Fig. 8(b) by the solid and dotted lines respectively. It can be seen that a spin dependent broadening $\Gamma_+ > \Gamma_-$ introduces an asymmetry in the spin-split FFT peak consistent with that observed experimentally (see Fig. 6 inset) demonstrating the validity of this interpretation.

This provides compelling insight to the nature of the spin-dependent scattering in these samples. Based on these simulations we can determine that *the low-energy (majority) spin state is the spin up state which undergoes greater scattering events than the high-energy (minority) spin down state* i.e. $\tau_+^{-1} > \tau_-^{-1}$. The uniqueness of this interpretation is demonstrated by reversing the asymmetry of the input broadening parameters i.e. $\Gamma_- > \Gamma_+$ with $\Gamma_+ = 1.4$meV and $\Gamma_- = 1.6$meV. The resulting FFT spectrum from this simulation is shown in Fig. 8(c), which clearly exhibits the opposite peak asymmetry. The assignment of the relative spin energies is as expected for a system with a negative g-factor. However, it is interesting to note that Lou *et al.* found the opposite in the InAs QW i.e. the low-energy spin state has a longer scattering time than the high-energy spin state [30]. From our analysis we can not determine whether the spin-dependent scattering rates originate from a spin-dependent scattering mechanism or simply from the differing densities associated with each spin population related to self-screening or even many body effects. There are no intentional magnetic materials incorporated during the growth of our heterostructures which may preferentially scatter one spin orientation more than another. In addition, we can not rule out the possibility that the large SO coupling in this system could play an important role in the scattering processes. Although our observations and those of Ref [30] differ, we note that the samples studied in Ref [30] are also structurally and electrically very different compared to ours; narrow QWs with high carrier densities and relatively low mobilities. It is feasible that one or more of the possible explanations for spin dependent scattering rates could be strongly influenced by these parameters.

## IV. CONCLUSIONS

In summary we have presented high field magneto-transport data from a range of high mobility InSb QW samples as a function of temperature and gate bias. A detailed analysis of the level broadening in these samples was made indicating a clear relationship with the carrier density. With the use of a top gate electrode we were able to modulate the carrier density in the 2DEG and detect beating in a number of samples with narrow broadening. Rashba coupling parameters were extracted from the difference in spin population determined from FFT analysis ranging from 1.3-1.5x10$^{-11}$eVm. With the use of numerical simulations of the SdH oscillations demonstrated that the absence of beating in many of the samples can be attributed to the combination of large inhomogeneous broadening combined with the presence of a spin dependent scattering rate; a phenomenon which has been shown to be manifested in systems with large spin splitting. The low-energy (majority) spin state is found to be the spin up state, consistent with the presence of a negative *g*-factor, which has a greater relaxation rate than the high-energy (minority) spin down state. This observation is counter to that found in the InAs system. The investigations of weak anti-localisation in the extreme low field limit is the subject of future work.

## ACKNOWLEDGEMENTS

This work was supported by the EPSRC. One of the authors (A.M.G) wishes to thank J. J. Harris for helpful discussions during the course of this work.

___________________________________________